\documentclass[12pt]{article}
\usepackage{a4}

\usepackage{amsfonts}
\usepackage{amsmath}
\usepackage{amssymb}
\usepackage{graphicx,color}
\usepackage{hyperref}

\newcommand{\bea}{\begin{eqnarray}}
\newcommand{\eea}{\end{eqnarray}}
\newcommand{\be}{\begin{equation}}
\newcommand{\ee}{\end{equation}}
\newcommand{\nn}{\nonumber}
\newcommand{\pa}{\partial}

\newcommand{\Ord}{{\cal O}}
\newcommand{\cP}{{\cal P}}

\newcommand{\mc}{m_{\sigma}}
\newcommand{\mad}{m_{AD}}
\newcommand{\mgt}{m_{3/2}}
\newcommand{\mh}{m_{1/2}}
\newcommand{\Vc}{V_\sigma}
\newcommand{\ts}{\tilde{\sigma}}
\newcommand{\bk}{{\bf k}}

\renewcommand{\Im}{\mbox{Im}}

\makeatletter

\@addtoreset{equation}{section}
\makeatother

\def\href#1#2{#2}

\begin{document}

\begin{titlepage}

\begin{center}

\hfill

\vskip 1.4in

{\large \bf Affleck-Dine baryogenesis
in inflating curvaton scenario}\\[2mm]
{\large \bf with $\Ord (10-10^2\, \mbox{TeV})$ mass moduli curvaton}\\[16mm]

{Kazuyuki Furuuchi${}^1$ and Chia-Min Lin${}^2$}
\vskip 10mm
{\sl ${}^1$National Center for Theoretical Sciences\\
National Tsing-Hua University, Hsinchu 30013, Taiwan R.O.C.}\\
{\tt furuuchi@phys.cts.nthu.edu.tw}\\[3mm]
{\sl ${}^2$Institute of Physics, Academia Sinica, Taipei 11529, Taiwan R.O.C.}\\
{\tt cmlin@phys.sinica.edu.tw}

\end{center}
\vspace{11pt}
\begin{abstract}
We study the Affleck-Dine (AD) baryogenesis
in the inflating curvaton scenario,
when the curvaton 
is a moduli field with
$\Ord (10-10^2\, \mbox{TeV})$ mass.
A moduli field with such mass
is known to be free from the Polonyi problem, 
and furthermore its decay products
can explain the present cold dark matter abundance. 
In our scenario, it further explains the
primordial curvature perturbation
and the present baryon density all together.
The current observational bound
on the baryon isocurvature perturbation,
which severely 
constrains the AD baryogenesis
with
the original oscillating moduli curvaton scenario,
is shown to put
practically negligible constraint
if we replace the oscillating curvaton
with the inflating curvaton.
\end{abstract}

\end{titlepage}

\setcounter{footnote}{0}

\section{Introduction}

The existence of moduli fields, 
light neutral scalar fields 
which interact with other fields
only with the strength of gravitational interaction,
is ubiquitous in
candidates of more fundamental theory like 
supergravity or string theory.
Since they interact only with the strength of gravity,
their life-time is long and their late decay might destroy the
successful scenario of the big-bang nucleosynthesis (BBN).
This is the so-called Polonyi problem
or cosmological moduli problem
\cite{Coughlan:1983ci,Ellis:1986zt,Banks:1993en,deCarlos:1993jw,%
Kawasaki:1999na,Kawasaki:2000en}.
The problem can be circumvented
if the mass of the lightest moduli field
is larger than some $10$ TeV scale
so that the reheating temperature is 
higher than $\sim 1$ MeV 
at which BBN starts \cite{Ellis:1986zt}.
Then, if the origin of the baryon number
in the universe is the Affleck-Dine (AD) mechanism 
\cite{Affleck:1984fy},
the dilution of the baryon number
by the decay of the moduli field may 
correctly explain the baryon asymmetry of the universe
\cite{Moroi:1994rs}.
Furthermore, if the non-thermal decay of the moduli field
produces the wino-like lightest supersymmetric particles (LSPs),
it may account for the cold dark matter abundance in the present universe
\cite{Moroi:1999zb}.
Thus, a moduli field with 
$\Ord (10-10^2 \,\mbox{TeV})$ mass
combined with the AD baryogenesis
may explain the baryon asymmetry and the dark matter abundance
of the universe all together.
Recently, it has been argued 
that string compactifications to four dimensions
with stabilized moduli may generically
have at least one moduli field
with $\Ord (10-10^2 \,\mbox{TeV})$ mass,
and the above scenarios have been revisited from this view point
\cite{Acharya:2008bk,Acharya:2009zt,Acharya:2010af,Kane:2011ih}.%
\footnote{%
The interests to the non-thermally produced
wino-like dark matter was also boosted
by the PAMELA data
\cite{Grajek:2008pg,Kane:2009if,Feldman:2009wv,Hisano:2008ti}.}
It will be very interesting if $\Ord (10-10^2 \,\mbox{TeV})$ mass
moduli field can further explain other features of our universe.

In this paper,
we examine whether $\Ord (10-10^2 \,\mbox{TeV})$ mass
moduli field
can explain the primordial curvature perturbation
as a curvaton \cite{Enqvist:2001zp,Lyth:2001nq,Moroi:2001ct,Lyth:2002my}.
We assume weak-scale supersymmetry 
with gravity-mediated supersymmetry breaking,
where the gravitino has 
$\Ord (10-10^2 \,\mbox{TeV})$ mass.\footnote{\label{foot2}%
The anomaly-mediation also works for
the non-thermal production of the cold dark matter,
but the AD baryogenesis part does not 
work in the same way as in the gravity-mediation case studied here.
See e.g. \cite{Kawasaki:2006yb} and references therein.}
In this case the moduli field also naturally has 
$\Ord (10-10^2 \,\mbox{TeV})$ mass.
Without special mechanisms, sfermions
also acquire the mass of similar scale.
However, if the field whose
F-term dominates the supersymmetry breaking
in the hidden sector is different from
the field whose vacuum expectation
value generates the gauge couplings,
gaugino masses can be suppressed 
\cite{Acharya:2007rc,Acharya:2008bk,Acharya:2009zt,%
Acharya:2010af,Feldman:2011ud,Kane:2011ih}.
We will be interested in the case where
the LSP is wino-like,
in order to explain the present cold dark matter abundance.
The AD baryogenesis in the original 
oscillating curvaton scenario
has been studied in \cite{Ikegami:2004ve}
where it was found that
the observational bound on the
baryon isocurvature perturbation
puts an important
constraint.
The constraint is severe
when the curvaton is a moduli field,
since there are natural mass scales for the moduli field
and the AD scalar field in weak-scale supersymmetry scenarios.
Interestingly,
this constraint naturally leads us to 
consider the recently proposed
inflating curvaton scenario \cite{Dimopoulos:2011gb},
as we explain in section \ref{secissue}.
Our scenario for the baryon asymmetry of the universe has
an important difference from the above mentioned scenario,
and this difference is crucial for the suppression of the
baryon isocurvature perturbation.
Our scenario for the baryogenesis
can be regarded as AD baryogenesis with
a version of low-scale inflation 
with low reheating temperature
\cite{Randall:1995dj,Asaka:2001xq}.

Before entering the main body,
let us 
summarize our scenario
for the history of the early universe.
The relevant scales are:
the Hubble expansion rate of the first inflation $H_1$,
the decay rate of the inflaton $\Gamma_I$,
the mass of the inflaton $m_I$,
the Hubble expansion rate of the second inflation $H_2$,
the mass of the inflating curvaton $\mc$,
the mass of the AD scalar field $\mad$,
the decay rate of the curvaton $\Gamma_\sigma \sim \mc^3/M_P^2$.
Here, $M_P$ is the reduced Planck scale:
$M_P = (8\pi G)^{-1/2} \sim 2.4 \times 10^{18}$ GeV.
Our scenario assumes 
$H_1 > m_I > \Gamma_I > H_2 > \mc > \mad > \Gamma_\sigma$:
\begin{enumerate}
\item The first inflation with the Hubble expansion rate $H_1$.
      The curvaton acquires quantum fluctuation,
      which becomes classical at the horizon exit.
      The phase part of the AD scalar also 
      acquires quantum fluctuation if 
      the Hubble induced A-term is suppressed,
      and this 
      leads to baryon isocurvature perturbation.
\item When $H\sim m_I$ the first inflation ends and 
      the inflaton starts coherent oscillations.
\item After the time scale $t \sim \Gamma_I^{-1}$, 
      the inflaton decays and reheats
      the universe.
      Then the energy density of the universe is dominated by radiation.
\item As the universe expands,
      the radiation is diluted and later on
      the potential energy of the curvaton
      begins to dominate the energy density of the universe.
      This leads to the second inflation
      with the Hubble expansion rate $H_2$.
      The primordial curvature perturbation is produced
      around the time when the second inflation starts.
      The pre-existed radiation is 
      diluted away by the second inflation.
\item When $H \sim \mc$, the second inflation ends
      and the curvaton begins coherent oscillations.
      The energy density of the universe is dominated
      by the curvaton matter.
\item When $H \sim \mad$, the AD scalar field begins a spiral motion
      and produces the baryon number.
\item After $H$ becomes much smaller than $\mad$,
      the baryon number freezes.
      The AD scalar eventually decays 
      and the baryon number is
      transmitted to the Standard Model (SM) particles.
      The energy density of the AD scalar 
      is always sub-dominant and its decay does not reheat
      the universe.
\item At $t\sim \Gamma_\sigma^{-1}$,
      the curvaton decays and reheats the universe.
      The cold dark matter is produced from 
      the decay product of the curvaton.
\end{enumerate}
The organization of this paper is as follows.
In section \ref{secinf}, we discuss
the inflating curvaton scenario,
which corresponds to the epoch 1-5 above.
In section \ref{secAD} we discuss the
AD baryogenesis in our scenario,
which corresponds to the epoch 5-8.
In section \ref{seciso}
we discuss the baryon isocurvature perturbation.
In section \ref{secissue}, we review
an issue in the AD baryogenesis in original 
oscillating curvaton scenario
for a comparison.
We end with summary and discussions in section \ref{secsummary}.
{
In appendix \ref{icq} we review
the argument of \cite{Dimopoulos:2011gb} 
for why and under what conditions
the inflating curvaton with
a quadratic potential cannot dominate
the primordial curvature perturbation.}
In appendix \ref{seccdm} 
we review the non-thermal production of
the wino-like cold dark matter
from the inflating curvaton decay,
which corresponds to the epoch 8 above.

\section{Inflating curvaton scenario}\label{secinf}

An interesting new scenario 
for the origin of the primordial 
curvature perturbation,
named ``inflating curvaton," was recently proposed
in Ref.\cite{Dimopoulos:2011gb}.
This scenario is radically 
different from the original oscillating curvaton scenario
\cite{Enqvist:2001zp,Lyth:2001nq,Moroi:2001ct,Lyth:2002my}.
As briefly reviewed in the introduction,
in this scenario the curvaton 
starts to dominate the energy density
while it is still slowly varying,
giving a few e-folds of inflation before it starts to oscillate.
In section \ref{secissue}
we will argue
that the inflating curvaton scenario 
is quite natural
when combined with the AD baryogenesis,
considering the observational bound on 
baryon isocurvature perturbation.
In the inflating curvaton scenario,
cosmological scales are demanded to be
outside the horizon
at the time $t_2$ when the second inflation starts.
Requiring that the shortest cosmological scale,
say a region enclosing $10^4$ solar masses, to be outside the horizon
at the beginning of the second inflation,
the e-folds $N_2$ required for the second inflation 
is given as \cite{Dimopoulos:2011gb,Lyth:2009zz}
\bea
 \label{N2}
N_2 \lesssim 45 
- \frac{1}{2} 
\ln 
\left(
\frac{10^{-5} M_P}{H_2}
\right).
\eea
In this paper we consider 
the pseudo-Nambu-Goldstone boson (PNGB) type 
inflating curvaton potential
considered in \cite{Dimopoulos:2011gb}:
\bea
 \label{cpot}
\Vc(\sigma)
=
\mc^2 f^2
\left[
1-\cos \left( \frac{\sigma}{f} \right)
\right].
\eea
For a moduli field,
$\mc \sim \mgt$ and $f \sim M_P$ would be
quite natural
and we will assume it to be the case in the following.
We will discuss a constraint on $f$
from the observations 
in section \ref{seciso}.
With the above parameter region, 
$H_2 \sim \mc$ and
thus the slow-roll condition is not satisfied;
we are in the regime of 
the fast-roll inflation \cite{Linde:2001ae}.
When
$\pi  - \sigma/f \ll 1$, 
the potential
(\ref{cpot}) can be approximated as
\bea
\Vc(\sigma)
\sim
\mc^2 f^2
-
\frac{\mc^2}{2}
\left(\pi f - \sigma \right)^2.
\eea
When $H$ is approximately constant:
\bea
 \label{H2}
H \sim H_2 = \frac{f}{\sqrt{6} M_P} \mc,
\eea
the evolution of the curvaton field
is described by the equation
\bea
 \label{eqc}
\ddot{\ts} + 3 H_2 \dot{\ts} 
- 
\mc^2 
\ts 
 = 0,
\eea
where 
\bea
\ts \equiv 
{\pi f}- \sigma .
\eea
In the above, the dot denotes the derivative with respect to time $t$
in the coordinate system 
\bea
 \label{FRW}
ds^2 = -dt^2 + a^2(t) \sum_{i=1}^3 (dx^i)^2.
\eea
(\ref{eqc}) can be solved by an ansatz
\bea
 \label{anz}
\ts = 
\left(
\pi f - \sigma_2
\right) 
e^{\omega (t-t_2)},
\eea
where we have set the initial condition when the 
second inflation starts as
$\sigma = \sigma_2 < \pi f$ at $t=t_2$.
The inflating curvaton rolls towards the origin $\sigma = 0$.
Putting (\ref{anz}) into (\ref{eqc}),
we obtain
\bea
\omega^2 + 3 H_2 \omega - \mc^2 =0,
\eea
thus
\bea
\omega = \omega_\pm \equiv \frac{-3H_2 \pm \sqrt{9H_2^2+4\mc^2}}{2} .
\eea
The solution with $\omega=\omega_-$ corresponds 
to the exponentially
decreasing field which rapidly disappears,
whereas the solution with $\omega = \omega_+$
corresponds to the exponentially growing field.
Below we consider the $\omega = \omega_+$ solution:
\bea
\ts
= \left(\pi f- \sigma_2\right) e^{F H_2 (t-t_2)} ,
\eea
where
\bea
F \equiv \frac{3}{2} 
\left(
\sqrt{1+\frac{4\mc^2}{9H_2^2}} -1
\right).
\eea
The field value of the curvaton
at the end of the second inflation 
$\sigma(t_e) \equiv \sigma_{e}$
may be well approximated by
$\sigma_{e} \sim \pi f/2$. 
Then the number of e-folds by the 
second inflation is given by
\bea
e^{N_2}
=
e^{H (t_e -t_2)}
\sim
\left(
\frac{\frac{\pi f}{2}}{\pi f - \sigma_2}
\right)^{1/F},
\eea
thus
\bea
 \label{N2r}
N_2 \sim \frac{1}{F} 
\ln 
\left(
\frac{\frac{\pi f}{2}}{{\pi f} - \sigma_2}
\right) .
\eea
When
$\sigma \ll \pi f/2$,
the potential (\ref{cpot}) can be 
approximated as
\bea
 \label{aqpot}
\Vc(\sigma)
\sim
\frac{\mc^2}{2} \sigma^2.
\eea
The curvaton starts to oscillate 
when 
$\sigma \lesssim \pi f /2$
and $H \sim m_\sigma$.

{
Now let us discuss the primordial curvature perturbation.
To the first order in the fluctuation of 
the energy density $\delta \rho$,
the primordial curvature perturbation $\zeta(\bk,t)$ is given as
\bea
\zeta(\bk,t) =
-H(t)\frac{\delta \rho(\bk,t)}{\dot{\rho}(t)}
=
\frac{1}{3}
\frac{\delta \rho(\bk,t)}{\rho(t)+ p(t)},
\eea
where $\rho$ is the energy density and
$p$ is the pressure.
The second equality corresponds to the
energy continuity condition $\dot{\rho} = -3H (\rho + p)$.
Just from the conservation of the energy-momentum tensor
one can show the conservation of the curvature perturbation
when the pressure is a unique function
of the energy density \cite{Lyth:2004gb,Lyth:2009zz}.
In any curvaton scenario, $\zeta$ is generated
while
$\rho = \rho_\sigma + \rho_r$
and
$p=p_\sigma + p_r$,
where 
$\rho_\sigma = \dot{\sigma}^2/2 + \Vc(\sigma)$
and 
$p_\sigma = \dot{\sigma}^2/2 - \Vc(\sigma)$ 
are the
curvaton contributions 
to the energy density and the pressure respectively,
and $\rho_r$ and $p_r$ are those from the radiation.\footnote{%
In inflating curvaton scenario those can be from the matter \cite{Dimopoulos:2011gb}.} 
Let us write
\bea
\zeta(\bk,t) = h(t)\zeta_\sigma (\bk,t),
\eea
where 
$h(t) \equiv (\rho_\sigma + p_\sigma) / (\rho+ p)$ and
$3\zeta_\sigma = \delta \rho_\sigma/(\rho_\sigma + p_\sigma)$.
There is supposed to be negligible
exchange of energy between the two components,
so that $\zeta_\sigma$ is constant
if $p_\sigma$ is a unique function of $\rho_\sigma$.
In inflating curvaton scenario,
$\zeta_\sigma$
becomes constant soon after 
the second inflation begins
\cite{Dimopoulos:2011gb}.
The second inflation dilutes the radiation and 
$h(t)$ soon becomes close to $1$.
Thus we have
\bea
\zeta(\bk)
\sim
\frac{\delta \rho_\sigma(\bk,t_2)}{3 \dot{\sigma}^2(t_2)}.
\eea
To the first order in $\delta \sigma$,
$\delta \rho_\sigma \sim \Vc'\delta \sigma$.
At the horizon exit during the inflation,
$\delta \sigma \sim H_1/2\pi$. 
Writing
$\tilde{\sigma}(t_2)$ as a function
of the field value 
of the curvaton
at the horizon exit $\tilde{\sigma}_*$
during the first inflation:
$\tilde{\sigma}(t_2)=g(\tilde{\sigma}_*)$, 
the primordial curvature perturbation 
is given as}
\bea
\cP^{1/2}_\zeta
&\sim&
\frac{g'}{3}
\frac{\Vc'}{\dot{\sigma}^2(t_2)}
\frac{H_1}{2\pi} .
\eea
Since (\ref{eqc}) is linear in $\tilde{\sigma}$,
$g(\tilde{\sigma})$ is also linear in $\tilde{\sigma}$. 
Putting $g(\tilde{\sigma}) \sim \tilde{\sigma}$ 
and the curvaton potential (\ref{cpot}),
we obtain
\bea
 \label{pc}
\cP^{1/2}_\zeta
&\sim&
\frac{1}{3}
\left(
\frac{\mc}{FH_2}
\right)^2
\frac{H_1}{2\pi \left({\pi f} - \sigma_2 \right)} .
\eea
This should be compared with the CMB normalization \cite{Komatsu:2010fb}
\bea
 \label{CMBN}
\cP_{\zeta}^{1/2} \sim 5 \times 10^{-5} .
\eea
We will come back to the comparison
with the CMB observation
in section \ref{seciso}.

\section{AD baryogenesis after the second inflation}\label{secAD}

In this section we study
the AD baryogenesis
taking place after the second inflation.
We assume the AD scalar originates
from the flat direction in the 
Minimal Supersymmetric Standard Model (MSSM)
\cite{Dine:1995uk,Dine:1995kz}.
The potential for the AD scalar $\phi$ is given by
\bea
 \label{ADpot}
V_{AD}(\phi)
=
(-cH^2+ \mad^2) |\phi|^2
+
\left(
\frac{A_H H + A \mgt}{M^{p-3}} \lambda \phi^p + h.c.
\right)
+|\lambda|^2 \frac{|\phi|^{2p-2}}{M^{2p-6}},
\eea
where $c,A$ and $\lambda$ are order one constants,
$H$ is the Hubble expansion rate,
and $M$ is the UV cut-off scale.
The sign in front of the quadratic part
must be negative
in order for the AD scalar to have large field value
when it starts the spiral motion,
which is required in the AD baryogenesis.
The coefficient of the Hubble induced A-term $A_H$
can be order one or can be much suppressed,
depending on the symmetry of the inflaton sector.
While its magnitude
does not affect the produced baryon number density,
it is of crucial relevance for 
the baryon isocurvature perturbation
\cite{Enqvist:1998pf,Enqvist:1999hv,Kawasaki:2001in,Kasuya:2008xp}.
We will discuss it in some detail in section \ref{seciso}.
We are interested in the case $A_H \ll 1$,
and consider this case in the following.

During the second inflation,
the AD scalar stays at the minimum of the potential (\ref{ADpot}).
We first assume
\bea
 \label{ADcond1}
c H_2^2 \gg \mad^2 .
\eea
From (\ref{H2}), for $f \sim M_P$, we have $H_2 \sim \mc$.
In section \ref{seciso} we will see that $f$ is 
bounded from above 
as $f \lesssim 5 M_P$
by the upper bound on
the e-folds of the second inflation, see (\ref{upbdf}).
We also assume that the effects of A-terms
to the minimum of the potential
of the AD scalar $\phi$
is smaller than those from the other terms
during the second inflation.
This assumption amounts to
\bea
\label{ADcond2}
4 c (p-1) H^2 \gg p^2 A^2 \mgt^2,
\eea
when $H\sim H_2$.
(\ref{ADcond1}) and (\ref{ADcond2})
may be satisfied with moderately large $c$.
For example, when $f = \sqrt{6} M_P$ and $H_2 = \mc \sim 2\mgt \sim 2 \mad$,
$A\sim 1$ and $p=9$
which are the parameter set we will use later,
(\ref{ADcond2}) reads
\bea
 \label{ADcond22}
c \gg 0.6 ,
\eea
which may be realized with 
the order one coefficient $c$.
With the same parameter set,
(\ref{ADcond1}) 
can be also satisfied.  
With the above assumptions,
the field value during the second inflation can be
estimated as
\bea
 \label{ADmin}
|\phi|
\sim
\phi_{min} (H)
\equiv
\left(
\frac{c}{|\lambda|^2} 
\right)^{\frac{1}{2(p-2)}}
M
\left( 
\frac{H}{\sqrt{p-1} M}
\right)^{\frac{1}{p-2}} .
\eea
The baryon number density $n_B$ is given by
\bea
n_B =  i q
\left(
\phi \dot{\phi}^* - \dot{\phi} \phi^* 
\right) ,
\eea
where $q$ is the baryon number
carried by the AD scalar $\phi$
and the dot denotes the derivative with respect to time $t$
in the coordinate system (\ref{FRW}).
The evolution of the baryon number density
follows from the equation of motion
of $\phi$ and is given by
\bea
 \label{bnev}
\dot{n}_B + 3H n_B =
2 q\,
\Im
\left[ 
\phi \frac{\pa V_A (\phi)}{\pa \phi} 
\right], 
\eea
where 
$V_A (\phi)$ is the A-term part of the AD scalar potential (\ref{ADpot}):
\bea
V_A (\phi) = 
\frac{A_H H + A \mgt}{M^{p-3}} \lambda \phi^p + h.c.
\eea
Since we are interested in the case $A_H \ll 1$,
we will neglect the Hubble induced A-term in the following.
Then, during the second inflation
the angular part of the complex scalar $\phi$
will take random value.
When $H \sim \mc$ and $\sigma \lesssim \pi f/2$,
the inflating curvaton starts to oscillate and
the energy density of the universe 
is dominated by the curvaton matter.
When $H \sim m_{AD}$, (\ref{ADcond2}) is 
no longer satisfied.
Then, the potential for the angular direction
from the A-term becomes relevant and
the AD scalar starts a spiral motion.
By multiplying $a^3(t)$ 
($a(t)$ is the scale factor in (\ref{FRW}))
to (\ref{bnev})
and integrating with respect to the time $t$,
we obtain
\bea
 \label{aqn}
a^3(t) n_B(t) 
&\sim&
q p
\int^{t_{sp}} dt'
a^3 (t') 
\frac{\mgt}{M^{p-3}}
\Im 
\left[ 
A  \phi^p
\right] \nn\\
&& +\, q p
\int_{t_{sp}}^t dt'
a^3 (t') 
\frac{\mgt}{M^{p-3}}
\Im
\left[ 
A  \phi^p
\right] .
\eea
The contribution from the second term of (\ref{aqn})
is small because 
of the two reasons:
(a) $\Im [A\phi^p]$ changes sign rapidly due to the 
spiral motion of the AD scalar $\phi$.
(b)
The amplitude of $|\phi|^p$ decreases with time as 
$H^{\frac{p}{p-2}} \propto t^{-\frac{p}{p-2}}$
in the matter dominant universe
and it also makes the contribution
from this term small.
Therefore, the baryon number density 
is produced dominantly
at the onset of the spiral move $t_{sp}$.
Since $a \propto t^{2/3}$ in the matter dominant universe,
the integrand of the first term of (\ref{aqn})
is proportional to $t^{\frac{p-4}{p-2}}$.
Then the integration gives
\bea
\label{nB0a}
n_B 
\sim
qp\,
{\mgt}
\left(
\frac{|\phi_{sp}|^p}{M^{p-3}}
\right)
2\sin[p \theta_{sp} + \arg(A)] \times \frac{p-2}{2p-6}t_{sp}.
\eea
Here, we set the initial condition for the 
baryon number density to be zero
at the end of the second inflation. 
From $a \propto t^{2/3}$ and
$H = \dot{a}/a \sim \frac{2}{3} t^{-1}$
in the matter dominant universe,
(\ref{nB0a}) becomes
\bea
\label{nB0}
n_B 
\sim
q
\frac{p(p-2)}{p-3}
{\mgt}
\left(
\frac{|\phi_{sp}|^p}{M^{p-3}}
\right)
\sin[p \theta_{sp} + \arg(A)]
\times \frac{2}{3\mad} ,
\eea
where we have parametrized the phase part of 
the AD scalar 
by $\theta$: $\phi = |\phi| e^{i\theta}$,
and $\theta_{sp}$ 
is the value of $\theta$
when the AD scalar starts 
the spiral motion.
$|\phi_{sp}|$ is defined with (\ref{ADmin}) as
\bea
 \label{nB2}
|\phi_{sp}| \equiv \phi_{min}(H=\mad)
\sim
M
\left( 
\frac{\mad}{\sqrt{p-1} M}
\right)^{\frac{1}{p-2}} .
\eea
The ratio of the baryon density
to AD scalar
at this epoch is given by
\bea
 \label{nbnphi}
\left(
\frac{n_B}{n_\phi}
\right)_{sp}
\sim
\frac{n_B}{\mad |\phi_{sp}|^2}
\sim
\frac{2q p(p-2)}{3(p-3)\sqrt{p-1}}
\frac{
\mgt}{\mad} 
\sin[p \theta_{sp} + \arg(A)] .
\eea
When the Hubble expansion rate becomes much less than $\mad$,
the baryon number of the condensate is frozen,
and later be converted to the baryon asymmetry.
As we check later,
the energy density of the AD scalar 
is always sub-dominant 
and its decay does not reheat the universe. 
The curvaton decays around the time scale
$\Gamma_\sigma^{-1}$,
where
$\Gamma_\sigma \sim \mc^3/M_p^2$.
The baryon density at this epoch is given by
\bea
n_B (t \sim \Gamma_\sigma^{-1}) 
\sim
\mad \phi_{sp}^2 
\left(
\frac{n_B}{n_\phi}
\right)_{sp}
\left(
\frac{\Gamma_\sigma}{\mad}
\right)^2,
\eea
where the factor $\Gamma_\sigma/\mad$ comes from the expansion of the
universe in the matter dominance:
$
a \propto t^{2/3}
$. 
After the curvaton decay,
the curvaton energy is converted to 
the radiation
with reheating temperature $T_R$.
We make an approximation that 
the curvaton energy density is instantaneously
transferred to the energy density of radiation
at $t\sim \Gamma_\sigma^{-1}$.
At this time $H\sim \Gamma_\sigma$ and thus
\bea
 \label{rhor}
\rho_r
=
\frac{\pi^2 g_\ast(T_R)}{30} T_R^4 
\sim
3 \Gamma_\sigma^2 M_P^2 ,
\eea
where $g_\ast(T)$ 
is the effective number of the 
contribution
of the massless degrees of freedom
to the energy density at temperature $T$.
From (\ref{rhor}) we obtain
\bea
T_R \sim
\left(
\frac{90}{\pi^2 g_\ast(T_R)}
\right)^{1/4}
\sqrt{\Gamma_\sigma M_P}.
\eea
The entropy density at this epoch is estimated as 
\bea
s = \frac{2\pi^2}{45} g_\ast(T_R) T_R^3 . 
\eea
The baryon number to entropy ratio 
at the time of the reheating
is given by
\bea
 \label{bent}
\frac{n_B}{s}
&\sim&
\frac{45}{2\pi^2 g_\ast(T_R) T_R^3}\mad \phi_{sp}^2 
\left(\frac{\Gamma_\sigma}{\mad}\right)^2
\left(
\frac{n_B}{n_\phi}
\right)_{sp} \nn \\
&\sim&
\frac{45}{2\pi^2 g_\ast(T_R)}
\frac{1}{\sqrt{p-1}}
\frac{M}{M_P} 
\left(
\frac{\mc}{M_p}
\right)^{\frac{3}{2}}
\left(
\frac{\sqrt{p-1} M}{\mad}
\right)^{\frac{p-4}{p-2}}
\left(
\frac{n_B}{n_\phi}
\right)_{sp} 
\nn \\
&\sim&
\frac{45}{2\pi^2 g_\ast(T_R)}
\frac{1}{\sqrt{p-1}}
\frac{T_R M}{M_P^2}
\left(
\frac{\sqrt{p-1} M}{\mad}
\right)^{\frac{p-4}{p-2}}
\left(
\frac{n_B}{n_\phi}
\right)_{sp}.
\eea
In our scenario there is no further
entropy production 
at later stage and 
the baryon to entropy ratio 
(\ref{bent})
is fixed until today.
To estimate (\ref{bent}),
let us be little bit more precise about the
decay rate $\Gamma_\sigma$.
The precise number is model dependent 
and here we choose
\bea
 \label{exdecay}
\Gamma_\sigma = 4 \,\frac{\mc^3}{M_P^2},
\eea
following \cite{Kane:2011ih}, 
as an representative value.
Then, 
for $\mc \sim 150$ TeV
we have $T_R \sim 70$ MeV, where $g_\ast (T_R) = 10.75$ is used.
For $M\sim M_P$, $p=9$ and $q\sim 1$,
(\ref{bent}) gives
\bea
 \label{bpn1}
\frac{n_B}{s}
\sim
6 \times 10^{-11}
\times
\left(
\frac{\mc}{150\, \mbox{TeV}}
\right)^{\frac{3}{2}}
\left(
\frac{75\, \mbox{TeV}}{\mad}
\right)^{\frac{5}{7}}
\left(
\frac{\mgt}{\mad}
\right).
\eea
Here, we have assumed $\sin[p\theta_{sp} +\arg(A)] \sim 1$.
(\ref{bpn1}) gives the
correct order of the
baryon to entropy ratio
of the present universe $9 \times 10^{-11}$
for the representative parameter set.
We need relatively 
large $p$ to account for the
present
baryon asymmetry of the universe.
This is because the reheating temperature,
which depends on the moduli curvaton mass
$\sim \Ord(10-10^2 \mbox{TeV})$,
is relatively low in our scenario.
The choice
$p=9$ has a good reason,
since larger $p$ gives
larger $\phi_{sp}$ and thus
gives larger contribution to the baryon number,
and $p=9$ is the maximal $p$
in MSSM 
\cite{Gherghetta:1995dv}. 

In the above we assumed that the
energy density of the AD scalar is always sub-dominant.
This can be easily checked to be the case.
When the AD scalar starts to oscillate,
the energy density of the universe is given by
\bea
\rho_{tot} (H\sim \mad) \sim 3 \mad^2 M_P^2.
\eea
On the other hand, the energy density of the 
AD scalar is given by
\bea
\rho_{AD} \sim 
\mad^2 \phi_{sp}^2
&\sim&
\mad^2 M^2 
\left(
\frac{\mad}{\sqrt{p-1}M}
\right)^{\frac{2}{p-2}} \nn\\
&\ll& \rho_{tot} (H \sim \mad),
\eea
where the last line is satisfied
when $\mad$ is sufficiently smaller than $M$.

In the analysis so far,
we were assuming that the AD condensate
evolves homogeneously after it formed.
In general, there is a possibility that
the AD condensate becomes unstable
with respect to spacial perturbations
and turns into non-topological solitons called
Q-balls 
\cite{Coleman:1985ki,%
Kusenko:1997zq,Kusenko:1997ad,
Enqvist:1997si,Enqvist:1998en}.
If Q-balls are formed, 
our scenario for the 
evolution of the universe 
may need to be modified.
However, while we have not made detailed analysis,
it seems likely that
Q-balls are not formed in our preferred parameter region
$\mad \gg \mh$,
where $\mh$ is the mass scale for the gauginos
\cite{Allahverdi:2005rh}.
In order for the Q-balls to be formed,
it is necessary that the potential for the 
AD scalar is flatter than $|\phi|^2$ at large field values.
After taking account the one-loop correction,
the potential for the AD scalar looks like
\bea
V_{AD, 1-loop} (\phi)
\sim \mad^2 |\phi|^2
\left(
1 + K \ln \frac{ |\phi|^2}{M^2}
\right) + \ldots ,
\eea
where the coefficient $K$ is determined
from the renormalization group equations,
see e.g. \cite{Nilles:1983ge,Enqvist:1997si,Enqvist:2000gq}.
Loops containing gauginos make a negative contribution
proportional to $\mh^2$,
while loops containing sfermions make a positive contribution
proportional to $\mad^2$.
Thus when the spectrum is such that
the gauginos are much lighter than the sfermions i.e.
$\mad \gg \mh$,
which is the case of our interest,
$K$ is likely to be positive and thus Q-balls will not be formed.
More complete analysis of the 
Q-balls is beyond the scope of the
current paper and
is left to the future investigations.

\section{Baryon isocurvature perturbation}\label{seciso}

As we mentioned in section \ref{secAD},
while the baryon number density
is not affected much by the Hubble induced A-term,
the magnitude of the coefficient $A_H$ of the
Hubble-induced A-term is crucially
relevant for 
the uncorrelated baryon isocurvature perturbation
\cite{Enqvist:1998pf,Enqvist:1999hv,Kawasaki:2001in,Kasuya:2008xp}:
If there is no sizable Hubble-induced A-term,
the phase part of the AD scalar is effectively massless
during the first 
inflation and acquires quantum fluctuations,
which leads to 
uncorrelated baryon isocurvature perturbation.
Below we study this case,
i.e. $A_H \ll 1$ in eq.(\ref{ADpot}).
If the Hubble induced A-term is sizable,
there is no uncorrelated 
baryon isocurvature perturbation 
and the observational bound on it
does not put any constraint on our model.

Since the curvaton dominates the energy density
when the baryon number is generated, 
the baryon number to entropy density ratio 
(\ref{bent}) does not depend
on the curvaton field fluctuations.
Therefore,
no correlated baryon isocurvature perturbation is produced
in our scenario.
This is an important difference
from the AD baryogenesis 
in the original oscillating curvaton scenario,
as we discuss in section \ref{secissue}.

At the horizon exit during the first inflation,
the phase part of the AD scalar acquires 
fluctuation:
\bea
 \label{dth}
\delta \theta = \frac{H_1}{2\pi |\phi_1|} ,
\eea
where $\phi_1$ is the field value of the
AD scalar during the first inflation. 
It is given by 
$\phi_1 \equiv \phi_{min}(H_1)$, where
$\phi_{min}(H)$ is given in (\ref{ADmin}).
The baryon isocurvature perturbation $S_{B}$ is defined as
\bea
 \label{iso}
S_{B}
\equiv
\frac{\delta \rho_B}{\rho_B}
-
\frac{3}{4}
\frac{\delta \rho_\gamma}{\rho_\gamma}
=
\delta 
\log 
\left( 
\frac{n_B}{s}
\right).
\eea
Here, $\rho_B$ and $\rho_\gamma$
are the present energy density of
baryons and photons, respectively.
We have used $\rho_\gamma^{3/4} \propto s$
in the above.
Substituting (\ref{bent}) (with (\ref{nbnphi}))
into (\ref{iso}),
we obtain
\bea
S_{B} \sim p \cot [p\theta_{sp} + \arg(A)] \delta\theta .
\eea
We assume that the curvature perturbation
is dominantly from the curvaton.
Then, from (\ref{dth})
we can estimate the uncorrelated
isocurvature perturbation
\bea
 \label{isoth}
|S_{B}^{(uncorr)} |
&\sim& p \frac{H_1}{2\pi |\phi_1|} \nn\\
&\sim&
\frac{p\sqrt{p-1}}{2\pi}
\left(
\frac{H_1}{\sqrt{p-1} M}
\right)^{\frac{p-3}{p-2}} .
\eea
In the first line we assumed 
$\cot [p\theta_{sp} + \arg(A)] \sim \Ord (1)$,
though one should keep in mind that 
the $|\cot|$ function can take any value between
$[0,\infty]$. 
The baryon isocurvature perturbation 
is constrained by
the current observational 
bound on the matter isocurvature perturbation \cite{Komatsu:2010fb}:
\bea
 \label{isoobs}
|S_{B}^{(uncorr)} | 
<
\frac{\Omega_c}{\Omega_B} \times \sqrt{\alpha_0} \, \cP_\zeta^{1/2}
\sim 7 \times 10^{-5} ,
\eea
where we have used $\alpha_0 < 0.08$.
Here, we have used
$\rho_B/\rho_c = \Omega_B/\Omega_c \sim 0.2$.
For $p=9$ and $M \sim M_P$,
by comparing (\ref{isoth}) and (\ref{isoobs})
we obtain an upper bound on $H_1$:
\bea
 \label{bH1}
H_1 \lesssim 
8 \times 10^{-6} M_P .
\eea
Let us study the implication of this bound
to the inflating curvaton model discussed 
in section \ref{secinf}.
Putting (\ref{bH1}) to (\ref{CMBN})
we obtain
\bea
 \label{bcfv}
\left(
{\pi f} -\sigma_2
\right)
=
\frac{1}{3}
\left(
\frac{\mc}{FH_2}
\right)^2
\frac{H_1}{2 \pi \cP_\zeta^{1/2}}
\lesssim \ts_c (f),
\eea
where 
\bea
\ts_c(f)
\equiv
\frac{1}{3}
\left(
\frac{\mc}{FH_2}
\right)^2
\frac{8\times 10^{-6}M_P}{2 \pi \cP_\zeta^{1/2}}.
\eea
We made clear that 
$\ts_c(f)$ depends on $f$ 
through $H_2$ and $F$.
Putting (\ref{bcfv}) to (\ref{N2r})
we obtain
\bea
 \label{N2lb}
N_2
\gtrsim
\frac{1}{F}
\ln
\left(
\frac{\frac{\pi f}{2}}{\ts_c(f)}
\right).
\eea
On the other hand,
since $H_2 \sim \mc$, from (\ref{N2})
we have $N_2 \lesssim 36$.
Thus (\ref{N2lb})
leads to an upper bound on $f$.
The right hand side of (\ref{N2lb})
is slightly complicated function of $f$,
but numerically solving it we obtained
\bea
 \label{upbdf}
f \lesssim 5 M_P  .
\eea
On the other hand,
for a successful AD baryogenesis
we require $H_2 > \mad$.
Then from (\ref{H2}) we obtain
\bea
f > \frac{\mad}{\mc} \sqrt{6} M_P .
\eea
In order that there is an allowed region for $f$
we need
\bea
\frac{\mad}{\mc} < \frac{5}{\sqrt{6}} .
\eea
This condition is satisfied
in our scenario,
since we assume $\mc > \mad$
as summarized in the introduction.

Going in the opposite direction,
if we set the ratio $\mad/\mc$,
we obtain a theoretical constraint on 
uncorrelated baryon isocurvature perturbation
in our model.
However, this bound terns out to be
very mild, practically giving no constraint:
For the case $\mad/\mc \sim 0.5$,
we numerically obtained the bound
\bea
\alpha_0 \gtrsim 10^{-25}.
\eea
This is an extremely mild constraint since the
lower bound would not be detectable
in a foreseeable future.
Note that this is just a lower bound, 
meaning the
the baryon isocurvature perturbation above the bound
might be detected in the future observation.

\section{An issue in the oscillating moduli curvaton scenario 
with AD baryogenesis}\label{secissue}

In this section
we review an issue in the AD baryogenesis
in the original 
oscillating curvaton scenario,
when the curvaton is a moduli field
\cite{Ikegami:2004ve}
{
(see also \cite{Gordon:2002gv})}.
Then we argue that this issue naturally
leads us to consider the AD baryogenesis in
the inflating curvaton scenario.
In this section we will use the same notation 
for the curvaton and related variables as before,
but notice that we are discussing a different scenario
in this section.
It will not cause any confusion 
if the readers keep in mind that
the oscillating curvaton scenario is discussed only in this section.

In the oscillating curvaton scenario,
it is assumed that 
when the curvaton starts to oscillate i.e. $H \sim \mc$,
the universe is dominated from the 
radiation whose energy density is
\bea
 \label{rro}
\rho_{r,o} \sim 3 \mc^2 M_P^2 ,
\eea
while the energy density of the curvaton is given by
\bea
 \label{rso}
\rho_{\sigma,o} \sim \frac{1}{2} \mc^2 \sigma_o^2 ,
\eea
where $\sigma_o$ is the field value of the curvaton at this moment.
Here, we consider the quadratic potential for the curvaton: 
\bea
 \label{qpot}
\Vc (\sigma) = \frac{1}{2} \mc^2 \sigma^2
\eea
which is most popular in the oscillating curvaton models,
instead of (\ref{cpot}).
We write the scale factor at this moment as $a_o$.
From (\ref{rro}) and (\ref{rso}),
by assuming that the radiation dominates the 
energy density of the universe at this epoch,
i.e. $\rho_{r,o} \gtrsim \rho_{\sigma,o}$,
we have
\bea
 \label{rdominant}
\sigma_o \lesssim \sqrt{6} M_P .
\eea
After this epoch, the energy density of the curvaton matter
decreases as
$\rho_\sigma (a) \propto a^{-3}$, 
whereas that of the radiation decreases as
$\rho_r (a) \propto a^{-4}$,
where $a$ is the scale factor (see (\ref{FRW})).
Thus the subsequent evolution is given as
\bea
\label{evolution}
\frac{\rho_\sigma (a)}{\rho_r (a)}
= 
\frac{a}{a_o} \frac{\rho_{\sigma,o}}{\rho_{r,o}}
=
\frac{a}{a_o} \frac{\sigma_o^2}{6 M_P^2}.
\eea
From (\ref{evolution}),
$a_{eq}$ when the energy density of the curvaton and 
that of the radiation become equal is given by
\bea
a_{eq} = a_o \frac{6 M_P^2}{\sigma_o^2} .
\eea
At this moment the energy density of the curvaton is given by
\bea
\rho_{\sigma} (a_{eq})
&=& 
\rho_{\sigma,o} \left(\frac{a_o}{a_{eq}}\right)^{3} \nn \\
&=&
\frac{1}{2} \mc^2 \sigma_o^2 
\left( \frac{6 M_P^2}{\sigma_o^2} \right)^{-3} \nn \\
&=&
\frac{1}{2} \mc^2 \left( \frac{\sigma_o^8}{(\sqrt{6} M_P)^6} \right) .
\eea
At the time of the radiation-curvaton equality,
the expansion rate $H_{eq}$ is obtained from
\bea
H_{eq}^2 
=
\frac{2}{3M_P^2} \rho_\sigma(a_{eq})
=
{2\mc^2} 
\left( \frac{\sigma_o}{ \sqrt{6} M_P} \right)^8,
\eea
thus
\bea
\label{Heq}
H_{eq}
=
\sqrt{2} \mc \left( \frac{\sigma_o}{\sqrt{6} M_P} \right)^4.
\eea
The correlated baryon isocurvature perturbation
crucially depends on whether
which of the following cases is realized \cite{Ikegami:2004ve}:
\begin{enumerate}
\item The AD field starts to oscillate 
when the radiation is dominant. 
\item The AD field starts to oscillate
when the curvaton is dominant. 
\end{enumerate}
Whether which case is realized depends on 
whether the mass of the AD field $\mad$ is bigger or smaller than
$H_{eq}$:
If
$H_{eq} \lesssim \mad$, the case 1 is realized while
$H_{eq} \gtrsim \mad$, the case 2 is realized.
Using (\ref{Heq}),
these conditions can be rewritten as
\bea
\label{cond1}
\sigma_o &\lesssim& \sqrt{6} M_P \left( \frac{\mad}{\sqrt{2} \mc}\right)^{1/4}: 
\mbox{case 1} \\
\label{cond2}
\sigma_o &\gtrsim& \sqrt{6} M_P \left( \frac{\mad}{\sqrt{2} \mc}\right)^{1/4}: 
\mbox{case 2} 
\eea
Let us first look at the case 1.
As we have seen in section \ref{secAD},
the baryon number is dominantly generated at
the time when the AD scalar starts the spiral motion,
$H \sim \mad$.
The curvaton number density at this epoch 
is 
$n_\sigma = \mc\sigma_{sp}{}^{2}/2$, 
where $\sigma_{sp}$ is
the curvaton field value at this epoch.
Thus the baryon to curvaton number density ratio
is proportional to $n_\sigma^{-1} \propto \sigma_{sp}{}^{-2}$.
In the case 1,
$\sigma_{sp} \propto \sigma_{o} \propto \sigma_\ast$,
where $\sigma_\ast$ is the field value of the curvaton
at the horizon exit.
The baryon to curvaton ratio
is converted into the baryon to entropy ratio
when the curvaton decays,
and it is fixed until today if there is no entropy production
at later time.
Thus
in this case we have a correlated isocurvature perturbation \cite{Ikegami:2004ve}
\bea
 \label{isocorr}
S_{B}^{(corr)} 
\sim \delta \ln ( \sigma_{\ast}^{-2} )
\sim -2 \frac{\delta\sigma_\ast}{\sigma_\ast} .
\eea
In the oscillating curvaton scenario with the 
quadratic potential (\ref{qpot}),
$\delta\sigma_\ast$ is given by
\bea
 \label{oscc}
\delta\sigma_\ast = \frac{H_1}{2\pi} .
\eea
Here, $H_1$ is the Hubble parameter at the inflationary stage.
Assuming that the primordial curvature perturbation
is dominantly produced by the curvaton,
$H_1$ is related to $\sigma_\ast$
through the CMB normalization:
\bea
 \label{Hcur}
\cP_\zeta^{1/2}
=
\frac{H_1}{3\pi \sigma_*}
= 5 \times 10^{-5}.
\eea
From (\ref{isocorr}), (\ref{oscc}) and (\ref{Hcur})
we obtain
\bea
|S_{B}^{(corr)}|
\sim
\left|
2 \frac{\delta\sigma_\ast}{\sigma_\ast} 
\right|
\sim
2\times 10^{-4}.
\eea
This is one order above 
the current observational bound 
\cite{Komatsu:2010fb}
\bea
|S_{B}^{(corr)}| \lesssim \frac{\Omega_c}{\Omega_B}\sqrt{\alpha_{-1}}\, \cP_\zeta^{1/2}
\lesssim 2 \times 10^{-5},
\eea
where we have used 
$\alpha_{-1} < 0.005$.
Thus the case 1 is excluded by the observation.

On the other hand,
in the case 2,
after the curvaton becomes dominant
in the energy density of the universe,
the Hubble parameter is determined 
by the curvaton field value, and vice versa.
Put it differently,
after the curvaton dominates the energy density,
in the gauge where
on each time slice 
the energy density is spatially uniform, 
the curvaton density is also spatially uniform.
In this gauge, the baryon number is
also produced uniformly.
Thus the baryon to curvaton ratio does not depend
on the fluctuation of the curvaton.
The baryon to curvaton ratio is later converted to 
baryon to entropy ratio when the curvaton decays,
thus there is no correlated isocurvature perturbation.

From (\ref{cond2}), 
the case 2
may be realized 
by the following two ways:
\bea
	&(i)& \sigma_o \gtrsim \sqrt{6} M_P \label{condi}  .\\
	&(ii)& \mc \gg \mad \label{condii} .
\eea
Due to the $1/4$ power  
dependence on the ratio $\mad/\mc$ in (\ref{cond2}),
in order to realize the case 2 
we need a large hierarchy between $\mc$ and $\mad$.
In the weak-scale supersymmetry framework, 
the anomaly mediation can give a hierarchy typically of order 
$\mad/\mc \sim 10^{-2} - 10^{-3}$ by one-loop factor.
{
If we consider the case of the maximal hierarchy 
$\mad/\mc \sim 10^{-3}$,
$(\mad/\mc)^{1/4} \sim 0.2$ and therefore
from (\ref{cond2}) and (\ref{rdominant}) we obtain
\bea
 \label{ll}
0.2  \lesssim \frac{\sigma_o}{\sqrt{6} M_P} \lesssim 1 .
\eea
%
We do not have an analytical control in 
the transition region 
between the case 1 (\ref{cond1}) and case 2 (\ref{cond2}),
which corresponds to the region around
$\sigma_o / (\sqrt{6} M_P) \sim (\mad/\mc)^{1/4}\sim 0.2$.
{Therefore, one may wonder whether
there is an window in the region (\ref{ll})
which is not ruled out by the observational bound
on the correlated baryon isocurvature perturbation.}
However, according to
the numerical analysis in \cite{Ikegami:2004ve}
with the updated WMAP data \cite{Komatsu:2010fb},
the region (\ref{ll})
seems to be 
ruled out.

We have seen that
even if we assume the maximal hierarchy between $\mc$ and $\mad$
$\mad/\mc \sim 10^{-3}$ which can be
naturally realized in the weak-scale supersymmetry scenario 
to achieve (\ref{condii}),
the oscillating curvaton scenario 
is severely constrained, if not 
ruled out, 
by the observational bound
on the baryon isocurvature perturbation. 
{
Actually,
as mentioned in the footnote \ref{foot2} and can be
understood from the analysis in section \ref{secAD},
with the mass scales natural
in the anomaly mediation
the AD baryogenesis 
does not work as in the gravity mediation case.
Thus we considered  
$\mc \sim 2 \mad$
as our reference in the previous sections,
which is natural in the gravity mediation.
This case is clearly ruled out 
by the observational bound
on the baryon isocurvature perturbation}.

While the lower bound in (\ref{ll})
is closely tied with 
the observations and is hard to avoid,
the upper bound is just a condition 
in order to stay in a particular scenario,
namely the oscillating curvaton scenario:
The upper bound in (\ref{ll})
came from the assumption 
that the radiation dominates 
the energy density of the universe
when the curvaton starts to oscillate, (\ref{rdominant}).
Thus this bound can be relaxed
if we assume instead that
the energy density of the curvaton
is comparable or larger than that of the radiation
at the time when the curvaton starts to oscillate.
However, in this case 
the curvaton energy density before its oscillation may
cause the second stage of inflation
and the scenario would need to be modified considerably.
This situation may better be studied
in the framework of 
the inflating curvaton scenario,
which is the main focus of the current paper.}\footnote{%
In \cite{Dimopoulos:2011gb}
it was argued 
that 
the inflating curvaton
with a quadratic potential 
cannot make a dominant contribution
to the primordial curvature perturbation,
when both the inflaton and the curvaton have canonical kinetic terms.
We review the outline of their arguments in appendix \ref{icq}.}

\section{Summary and discussions}\label{secsummary}

The main results of this paper can be summarized as follows:
\begin{enumerate}
 	\item The AD baryogenesis in the inflating curvaton scenario
        is consistent with the
 	      observational bound on baryon isocurvature perturbation.
        Note that as explained in section \ref{secissue},
        the observational bound 
        on correlated baryon isocurvature perturbation
        severely 
        constrains the AD baryogensis
        in the original oscillating curvaton scenario
        when the curvaton is a moduli field.
	\item The moduli field with $\Ord(10-10^2\, \mbox{TeV})$ mass
	      plays multiple key roles in our scenario.
        It explains
	      the primordial curvature perturbation
	      as well as
        the baryon density and
	      the cold dark matter density of the present universe
	      (see appendix \ref{seccdm} for the cold dark matter part).
\end{enumerate}

It will be interesting
to realize our scenario
in a controlled string compactification
with stabilized moduli,
which will predict more precise values
for the physical input parameters.
It will also be interesting to examine whether
such
$\Ord (10-10^2\,\mbox{TeV})$ mass moduli field
exists in a large class of four-dimensional compactifications of
string theory with stabilized moduli 
\cite{Acharya:2010af}.

\vskip7mm
\noindent
\centerline{\bf Acknowledgments}\vskip1mm
We would like to thank 
Kazunori~Kohri and Fuminobu~Takahashi for 
explaining their works. 
We also thank 
Chian-Shu~Chen for discussions.
KF was benefited from the NCTS topical program
``LHC Physics: W, Z and beyond" held in Oct.~18 -- Dec.~17 2010,
where he learned from Gordy~Kane about their works.
He thanks the organizers and the participants
for creating stimulating environments.
KF is supported in part by 
the NCTS String Theory Focus Group 
under the NSC grant No.100-2119-M-002-001.

\appendix

{

\section{Contribution of 
the inflating curvaton with a quadratic potential
to the primordial curvature perturbation}\label{icq}

In this appendix 
we 
outline 
the arguments
of \cite{Dimopoulos:2011gb} 
that
the inflating curvaton with
a quadratic potential
cannot dominate the
primordial curvature perturbation
when both the inflaton and the curvaton have
canonical kinetic terms.

We start from looking at
the tilt of the spectrum
\cite{Lyth:2001nq,Wands:2002bn,Lyth:2009zz}
\bea
\label{tilt}
n(k) - 1
\equiv
\frac{d \ln \cP_\zeta}{d \ln k} = -2 \epsilon_{H1} + 2 \eta_1 
-\frac{2}{M_P^2\cP_\zeta(k)}\left( \frac{H_1(k)}{2\pi}\right),
\eea
where
\bea
\epsilon_H \equiv 
\frac{\dot{H}}{H^2},
\qquad
\eta \equiv
\frac{1}{3H^2}\frac{\partial^2 V(I,\sigma, \ldots)}{\partial \sigma^2}.
\eea
Here, $V(I,\sigma, \ldots)$ is the total potential for the inflaton
$I$ of the first inflation, the curvaton and other scalar fields in the model.
The subscripts $1$ mean they are the values during the first inflation. 
The right hand side of (\ref{tilt})
is evaluated at the horizon exit $k = aH$.
The observation gives $n-1 \sim 0.04$ \cite{Komatsu:2010fb}.
In many models 
the last two terms 
in (\ref{tilt}) 
are negligible.
Even when they are not negligible,
it is unlikely that
the terms in the right hand side of
(\ref{tilt}) cancel accurately 
so that they give the value in the observed tilt. 
Thus we obtain 
\bea
\label{eH1}
\epsilon_{H1} \lesssim 0.02.
\eea
Assuming that the inflaton has a canonical kinetic term,
the contribution to the primordial curvature perturbation
from the inflaton is given by
\bea
\label{PI}
\cP_{\zeta_I}^{1/2} \sim \frac{1}{\sqrt{2\epsilon_{H1}}}\frac{H_1}{2\pi M_P}.
\eea
On the other hand, 
when the curvaton also has a canonical kinetic term,
in order to 
realize the second inflation
we should require the 
slow-roll condition
$\epsilon_{H2} \sim \epsilon_2 \equiv M_P^2 (V'/V)^2 \ll 1$.
In this case, 
the contribution 
to the primordial curvature perturbation
from the curvaton is given by
\bea
\label{Pc}
\cP_{\zeta_\sigma}^{1/2} \sim
\frac{g'}{\sqrt{2\epsilon_2}}\frac{H_1}{2\pi M_P}.
\eea
Thus we obtain the ratio
\bea
\label{ratio}
\frac{\cP_{\zeta_\sigma}}{\cP_{\zeta_I}}
\sim
(g')^2\frac{\epsilon_{H1}}{\epsilon_2}
\sim
2 N_2 \, \epsilon_{H1} \left(\frac{\sigma_2}{\sigma_*}\right)^2,
\eea
where we have used $\sigma_2 = g(\sigma_*) \propto \sigma_*$.
Since the curvaton rolls down slowly
during the first and the second inflation,
there should not be much difference between
$\sigma_2$ and $\sigma_\ast$, $\sigma_2 / \sigma_\ast \lesssim 1$.
Thus from (\ref{N2}) and (\ref{eH1}),
we obtain
\bea
\label{ration}
\frac{\cP_{\zeta_\sigma}}{\cP_{\zeta_I}} \lesssim 1.
\eea
From (\ref{ration})
we conclude that with the naturalness argument above
eq.(\ref{eH1}),
the inflating curvaton with a quadratic potential
cannot dominate the primordial curvature perturbation
when both the inflaton and the curvaton have canonical kinetic terms.

}

\section{Non-thermal production of the cold dark matter}\label{seccdm}

For completeness, 
in this appendix
we review the non-thermal production of
the cold dark matter density  
\cite{Moroi:1999zb}
(see also \cite{Kawasaki:2007yy,%
Acharya:2008bk,Acharya:2009zt,Acharya:2010af,Kane:2011ih,Feldman:1900zz})
and confirm that it is realized in our model.
The moduli field couples to
other fields universally with the strength of 
the gravitational interaction,
thus it decays to the superpartners with a large branching ratio.
Each of these superpartners eventually decay
to an LSP. 
When the branching ratio is order one,
the LSP to entropy ratio
is roughly the same order
with 
the curvaton to entropy ratio 
before the curvaton decay.\footnote{%
As can be seen from the discussion below,
the conclusion would not change up to the branching ratio
as small as $\Ord(10^{-4})$.}
It can be calculated in a similar way
to the baryon to entropy ratio (\ref{bent})
and is given by
\bea
\frac{n_\chi}{s}
\sim
\frac{n_\sigma}{s}
\sim
\frac{45}{2 \pi^2 g_\ast(T_R) T_R^3}
\frac{3\mad^2 M_P^2}{\mc}
\left( 
\frac{\Gamma_\sigma}{\mad}
\right)^2
\eea
Here, we have approximated the inflating curvaton potential
with the quadratic potential (\ref{aqpot})
when $H\sim \mad$.
The produced LSPs undergo an
out-of equilibrium annihilation
if the self-annihilation rate is larger
than the expansion rate:
$n_\chi \langle  v_{rel}\sigma \rangle > H$.
This amounts to the following condition:
\bea
n_\chi \gtrsim 
n_\chi^c 
\equiv 
\frac{H}{\langle  v_{rel}\sigma  \rangle}
\Biggr|_{T=T_R} .
\eea
For 
the mass of the LSP
$m_\chi \sim 100$ GeV 
the cross section of the wino of this mass  
$\langle  v_{rel} \sigma \rangle \sim 3 \times 10^{-7}$ GeV$^{-2}$
\cite{Moroi:1999zb}.
Then for $\mc \sim 150 \mbox{TeV}$
the abundance is too large, i.e.
$n_\chi / s \sim  10^{-7}$ 
while
$n_\chi^c /s \sim 10^{-12}$.
Thus the LSPs further annihilate.
The final abundance is determined by the 
critical number density $n_\chi^c$.
The final dark matter to entropy ratio is given by
\bea
 \label{me}
\frac{n_\chi^c}{s}
=
\frac{45}{2\pi^2g_\ast(T) T^3} 
\frac{H}{\langle  v_{rel} \sigma \rangle}
\biggr|_{T=T_R}
.
\eea
We will use $g_\ast (T_R)=10.75$ as before.
(\ref{me}) can be converted into the relic abundance today:
\bea
 \label{OCDM}
\Omega_{\chi}
&=&
\frac{m_\chi}{\rho_{0}/s_0} \frac{n_\chi^c}{s} \nn\\
&\sim&
0.1 \,  h^{-2}
\left(
\frac{m_\chi}{100\, \mbox{GeV}}
\right)
\left(
\frac{3\times 10^{-7}\, \mbox{GeV}^{-2}}{\langle v_{rel}\sigma \rangle}
\right)
\left(
\frac{150\, \mbox{TeV}}{\mc}
\right)^{3/2} .
\eea
Here, $\rho_0/s_0$
is the ratio between the critical density 
and the entropy density 
$\rho_0/s_0 \sim 3.6 
\times 10^{-9} h^2$ GeV,
where $h$ is the present Hubble constant in units of
$100$ km s${}^{-1}$ Mpc${}^{-1}$.
(\ref{OCDM}) gives the correct order for 
the present dark matter to critical density ratio
$\Omega_{c} h^{-2} = 0.11$
for the representative set of parameters.\footnote{%
For a given reheating temperature $T_R$,
$m_\chi$ can be adjusted to obtain the observed
$\Omega_\chi$ as can be seen from (\ref{me}).
But the point is that 
the wino-mass chosen in this way is
quite natural 
in the current weak-scale supersymmetry scenario.}

\bibliography{adicref}
\bibliographystyle{utphys}

\end{document}